\documentclass{aa}  
\usepackage{graphicx}
\usepackage{txfonts}

\begin{document} 

\title{Unravelling the Post-Collision Properties of the Cartwheel Galaxy: A MUSE Exploration of its Bar and Inner Region}
\subtitle{}

\author{Chayan Mondal
          \inst{1}\fnmsep\thanks{corresponding author}
          \and
          Sudhanshu Barway\inst{2}
          }

   \institute{Inter-University Centre for Astronomy and Astrophysics, Ganeshkhind, Post Bag 4, Pune 411007, India\\
              \email{chayanm@iucaa.in, mondalchayan1991@gmail.com}
         \and
             Indian Institute of Astrophysics, Koramangala II Block, Bangalore-560034, India\\
             }

   \date{}

\abstract
{}
{To investigate the characteristics of the bar and inner disk in the collisional ring galaxy Cartwheel.}{We used the Integral Field Unit (IFU) observations with Multi-Unit Spectroscopic Explorer (MUSE) of the Very Large Telescope (VLT) to investigate the stellar kinematics, age, and nature of ionised gas in the inner region of the Cartwheel. We produced the stellar line of sight (LOS) velocity (V), velocity dispersion ($\sigma$), h$_3$ velocity moment, stellar population age, and emission-line maps of the galaxy using the Galaxy IFU Spectroscopy Tool (GIST) pipeline.}{The observed nature of intensity, V, and $\sigma$ profiles altogether support the existence of a stellar bar as earlier revealed from near-infrared (NIR) $K_s$ band imaging. A weak correlation between V/$\sigma$ and h$_3$ is found within the bar radius, providing more kinematic evidence for a stellar bar which survived the drop-through collision. The overall weak anti-correlation between V/$\sigma$ and h$_3$ in the disk implies that the stellar orbits in the disk are less stable, which might be due to the impact of the collision. The mass-weighted age map of the galaxy shows that the stellar populations in the bar region are relatively older, with an increasing gradient from the bar edge to the centre, another evidence to signify that the bar was present before the galaxy underwent collision. We do not find an active galactic nuclei (AGN) from the BPT analysis of a central unresolved source reported earlier using NIR imaging. Our findings provide the preservation of the pre-collisional structures in the inner region of the Cartwheel, an important input to understanding the evolution of collisional galaxy systems, particularly investigating the pre-collisional central region for theoretical studies.}{}
 

\maketitle

\begin{figure*}
    \centering
    \includegraphics[width=6in]{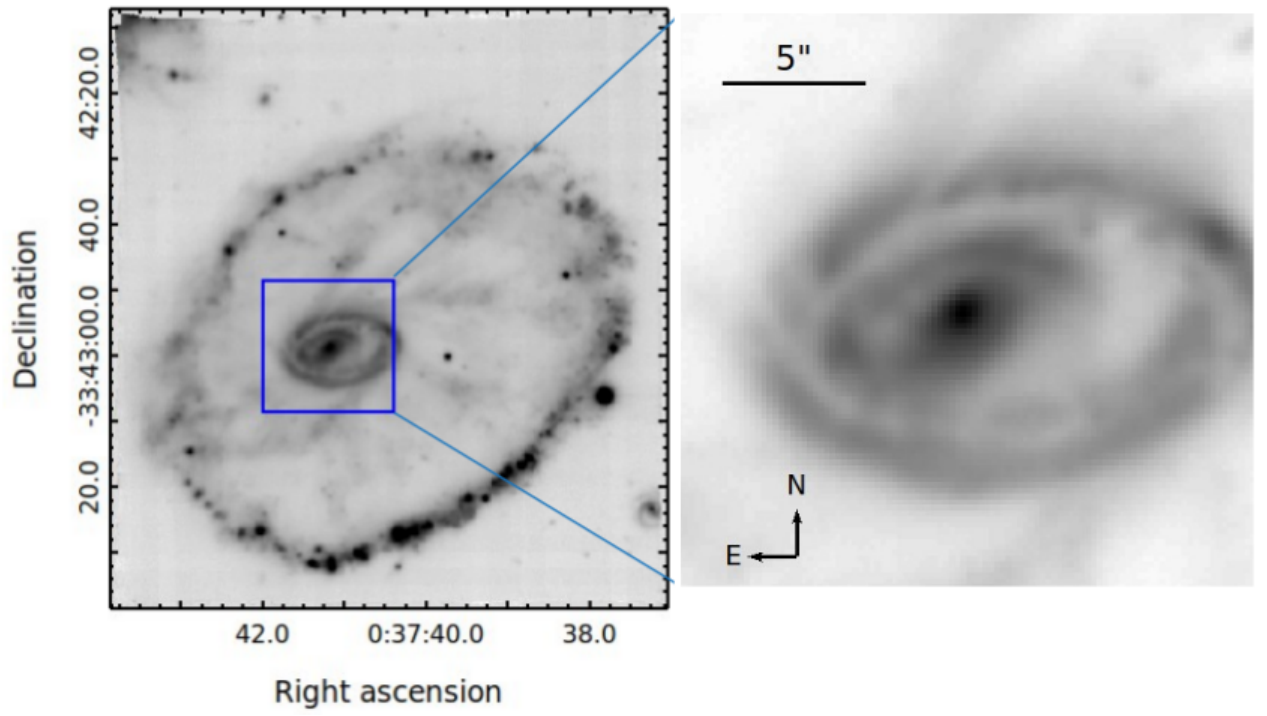} 
    \caption{The MUSE image of the Cartwheel galaxy. Left: the optical image of the Cartwheel ring galaxy extracted from the MUSE deep data cube for the wavelength range 4800 - 5800 ~\AA. Right: the zoomed-in view of the Cartwheel's inner region (i.e., the part studied in this work) as shown by the blue box of size 20$^{\prime\prime}\times$20$^{\prime\prime}$ on the left panel. The north is up, and the east is to the left.}
    \label{cartwheel}
\end{figure*}

\section{Introduction}
Stellar bars are extended linear structures frequently found in nearby disk galaxies, where they reshape bulges, regulate star formation, and drive the host galaxies' transformation \citep{kormendy2004,sakamoto1999,ellison2011,kim2016,kruk2018,dono2019}. Strong bars are found in more than 60\% of the disk galaxies that dominate star formation in the local universe \citep{eskridge2000,menendez2007,marinova2007}. The stellar bar is a long-lived and robust structure; however, it is unknown when the first bar formed in disk galaxies. Many simulations show that bars are strong structures and difficult to dissolve once formed \citep{athan2005,kraljic2012}. So, in the absence of legitimate pieces of evidence and clues, it remains a more complex problem to determine whether the same bar still exists in present-day galaxies or if it has been dissolved and rejuvenated \citep{bournaud2002,barway2020a}. 

The stellar bars are found across all environments and have a higher fraction in group/cluster environments, particularly for S0 galaxies \citep{barway2011}. The evolution of bar in such environments is poorly studied, particularly in the case when a galaxy undergoes an interaction and/or collision. Recently, \citet{barway2020} discovered a bar in the Cartwheel using Near-infrared (NIR) Ks-band images, making it the first collisionally formed ring galaxy to have a bar detected. The Cartwheel is the archetype of collisional ring galaxies and is formed by a compact galaxy falling head-on through a massive gas-rich disk galaxy close to its centre and almost perpendicular to it \citep{lynds1976}, and is distinguished by an outer ring that contains a chain of star-forming knots \citep{marston1995,appleton1997,higdon2015}. 

The Cartwheel has a distinct inner ring that, unlike the outer ring, is gas-poor, with little presence of H I \citep{higdon1996}, H$_{\alpha}$ \citep{amram1998}, and star formation \citep{charmandaris1997}. The inner ring's low star formation rate has cast doubt on its interpretation as a post-collisional structure, and it could equally well be of pre-collisional origin. Several studies have attempted to simulate Cartwheel's observed structure through a collision of two galaxies. Most of these have explored the nature of the outer ring and spokes in detail \citep{marcum1992, appleton1997, renaud2018}. There are very few studies through numerical simulations to understand the evolution of a pre-collisional bar or the inner disc's orientation after the collision. A theoretical study by \citet{athan1997} using N-body simulation to create the rings in galactic discs by infalling small companions attempts to address the questions mentioned earlier. These authors found that the bar survives the interaction and gradually grows fatter, eventually forming an oval structure. A small ring surrounds the bar and expands and detaches from it, forming an arm on one side. 

The studies using Multi-Unit Spectroscopic Explorer (MUSE; \citet{bacon2010}) observations at the Very Large Telescope (VLT) also focus primarily on the outer ring of the Cartwheel galaxy \citep{zaragoza2022, mayya2023}. In this paper, we present a detailed analysis of the inner region of the Cartwheel galaxy using MUSE data, which covers the entire optical extent of the galaxy at the seeing-limited angular resolution of $\sim$0\farcs6. Utilising the MUSE IFU data, we studied the kinematics and star formation history of the inner part to explore the existence of a pre-discovered bar in the Cartwheel. We also probed the nature of ionised gas and the gas-phase metallicity of the same part using identified emission lines in the MUSE data. Our primary aim is to provide more evidence for the presence of a bar in this collisional ring galaxy and also to understand the overall properties of its inner disk.

This article is organised as follows. Section \S\ref{s_bar_property} presents the photometric properties of the inner region. Section \S\ref{s_muse_data} describes the MUSE data set, the detailed analysis is presented in Section \S\ref{s_analysis}, and the results are presented in Section \S\ref{s_result}. We present our conclusions in Section \S\ref{s_summary}.  

\begin{figure*}
    \centering
    \includegraphics[width=6.5in]{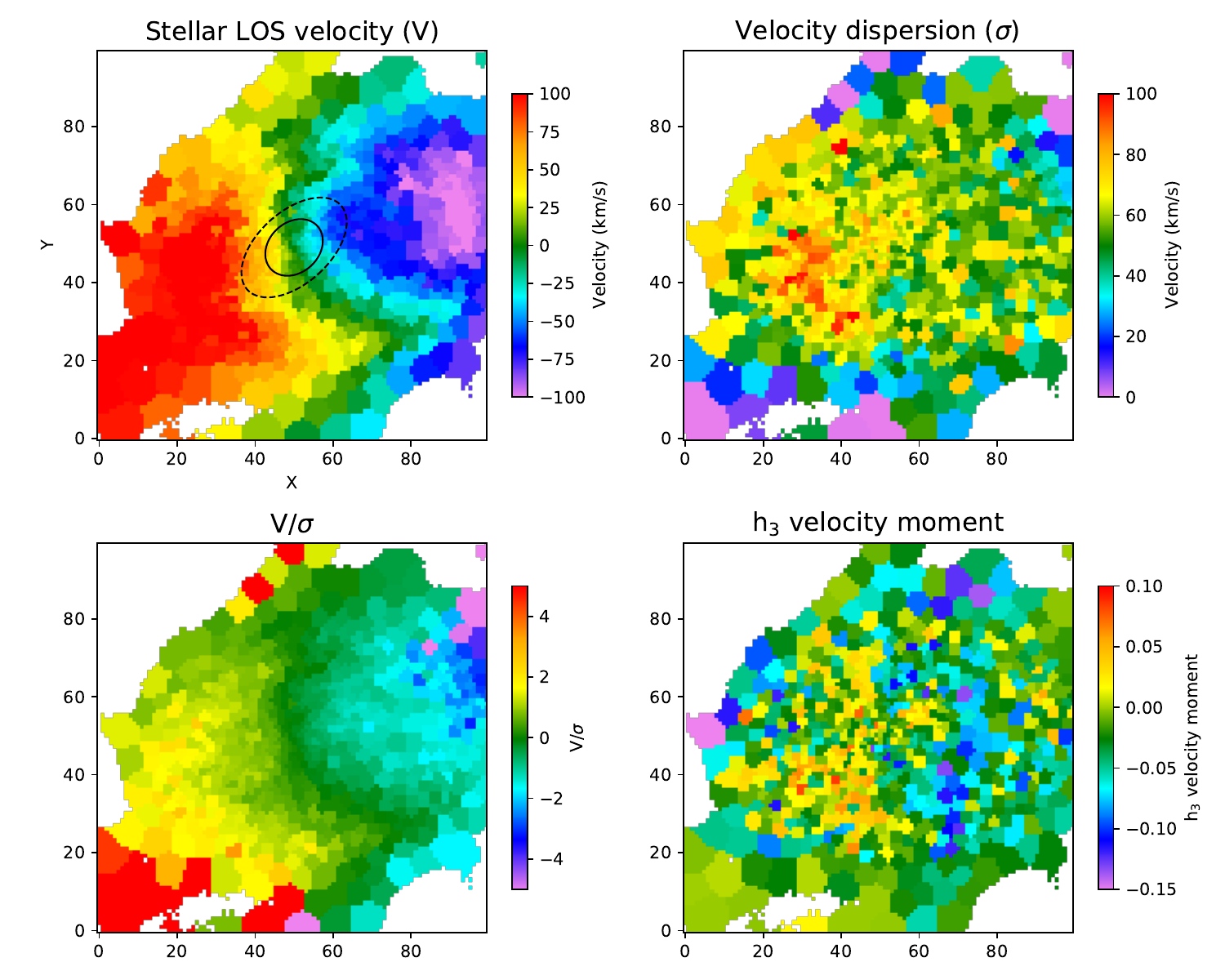} 
    \caption{Stellar kinematic maps of the Cartwheel. The stellar LOS velocity (V) distribution for the inner region (as shown by the blue rectangle in Figure \ref{cartwheel}) of the Cartwheel produced using GIST pipeline on the MUSE data cube (\textit{top-left}). Ellipses shown in dashed and solid black lines indicate the bar and bulge of the galaxy, respectively, as reported in \citet{barway2020}. The north is up, and the east is to the left. The stellar velocity dispersion ($\sigma$) (\textit{top-right}), V/$\sigma$ (\textit{bottom-left}), and the h$_3$ velocity moment (bottom-right) maps are also shown for the same part of the galaxy.}
    \label{v_by_sigma_bar}
\end{figure*}

\section{Photometric properties of Cartwheel bar}
\label{s_bar_property}
In general, stellar bars are predominantly composed of older stellar populations. As a result, NIR observations, known to reveal structures that are not always visible in optical images, are frequently used to identify bars in galaxies. The NIR emission from older stars is less affected by the dust extinction and the presence of younger stars \citep{cardelli1989,riad2010}. Utilising this property of the NIR emission, \citet{barway2020} reported the discovery of a stellar bar, a pseudo-bulge, and an unresolved point source in the archetype collisional ring galaxy Cartwheel through a careful morphological analysis of an excellent quality near-infrared $K_s$-band image from the European Southern Observatory (ESO) archive taken at the VLT-U1 telescope in the year 2000 using the Infrared Spectrometer And Array Camera (ISAAC) instrument. Due to a combination of its red colour and the presence of dusty features in the central region of the Cartwheel galaxy, the newly discovered bar is not recognisable in optical images even with a spatial resolution of the Hubble Space Telescope (HST).

Cartwheel's stellar bar is oval-shaped, with a semi-major axis length (bar length) of 2.09 kpc and a nearly flat light distribution. The bulge is nearly circular, with an effective radius of 1.05 kpc and a Sersic index of 0.99, which are typical of late-type galaxies' pseudo-bulges. Another important bar characteristic is the 'strength' of a bar, which is 0.36 for the Cartwheel bar as determined by \citet{barway2020}, a value typically found in late-type spiral galaxies with a strong bar \citep{aguerri2009}. To summarise, Cartwheel's stellar bar is a strong and robust structure that survived the collision and was unaffected by the impact. Collisional ring galaxies offer the opportunity to study a new, less explored aspect of bar evolution, and kinematic details using MUSE will be critical to investigate.

\section{MUSE IFS Data}
\label{s_muse_data}
We have used Integral Field Spectrograph (IFS) data of the Cartwheel observed with the MUSE instrument. MUSE is an optical IFS instrument installed on ESO's Very Large Telescope \citep{bacon2010}. The instrument operates within the wavelength range of 4750~\AA~to 9350~\AA~with 1$^{\prime}$ $\times$ 1$^{\prime}$ almost square field of view. The detector has the capability of 0\farcs2 spatial sampling and an average 1.25~\AA~ spectral sampling. The galaxy Cartwheel was observed between UT 24-08-2014 to 25-08-2014 as part of MUSE's science verification program (ID- 60.A-9333). We have obtained the science-ready MUSE deep data cube of the galaxy from the \textit{ESO archive science portal}\footnote{http://archive.eso.org/scienceportal/home} (archive ID - ADP.2017-03-23T15:47:52.027). The deep cube has a sky coverage of 2.2 arcmin$^2$ with a total integration time of 8400 sec. During the observation, the effective seeing was $\sim$0\farcs63. In Figure \ref{cartwheel}, we have shown the 2d intensity map of the galaxy, constructed from the obtained deep cube, for the wavelength range 4800 - 5800~\AA. The 2.2 square arcmin MUSE deep cube covers the entire galaxy, including the bright and distinct outer ring. In this work, we have studied only the inner $\sim$20$^{\prime\prime}\times$20$^{\prime\prime}$ part of the galaxy as shown on the right panel of Figure \ref{cartwheel}. We produced a smaller cutout of dimension 100$\times$100 spaxels for the inner part from the larger deep cube and used that for further analysis. At the distance of Cartwheel (which is 133 Mpc as per the \textit{NASA/IPAC Extragalactic Database}\footnote{https://ned.ipac.caltech.edu/}), each MUSE spaxel corresponds to a length scale of $\sim$130 pc in the galaxy frame. The MUSE intensity map of the galaxy's inner part clearly shows a bright central component surrounded by another ring structure. Our cutout encloses the inner ring including some parts of the spokes.

\begin{figure*}
    \centering
    \includegraphics[width=6.5in]{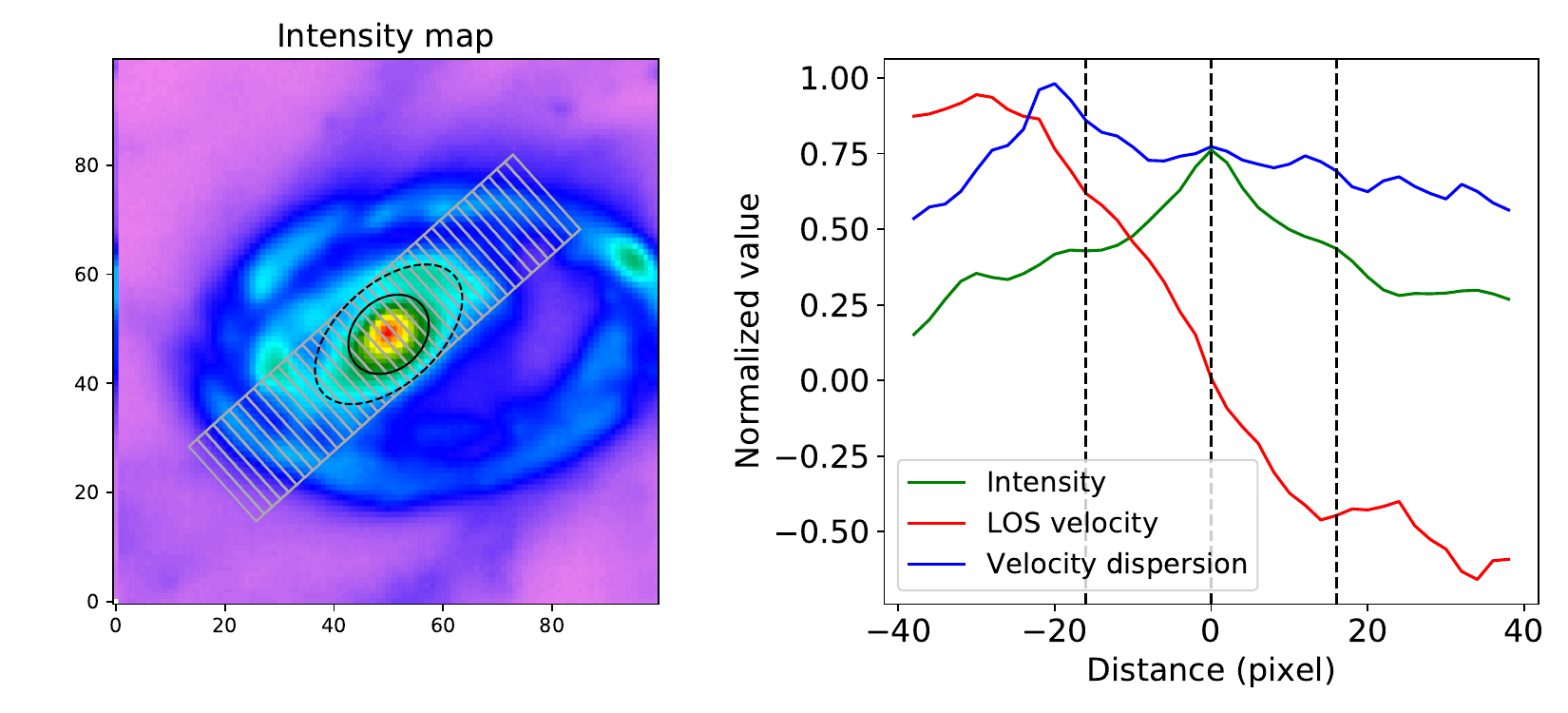} 
    \caption{\textbf{Left:} The intensity map for the inner region of the Cartwheel. The extent of the bar and bulge are shown by ellipses same as Figure \ref{v_by_sigma_bar}. \textbf{Right:} Intensity, LOS velocity, and velocity dispersion profiles along the grey rectangular pseudo-slit shown on the left panel. The vertical black dashed line at the centre signifies the position of the minor axis whereas those on either side highlight the bar end.}
    \label{all_profile_bar}
\end{figure*}

\begin{figure}
    \centering
    \includegraphics[width=3.5in]{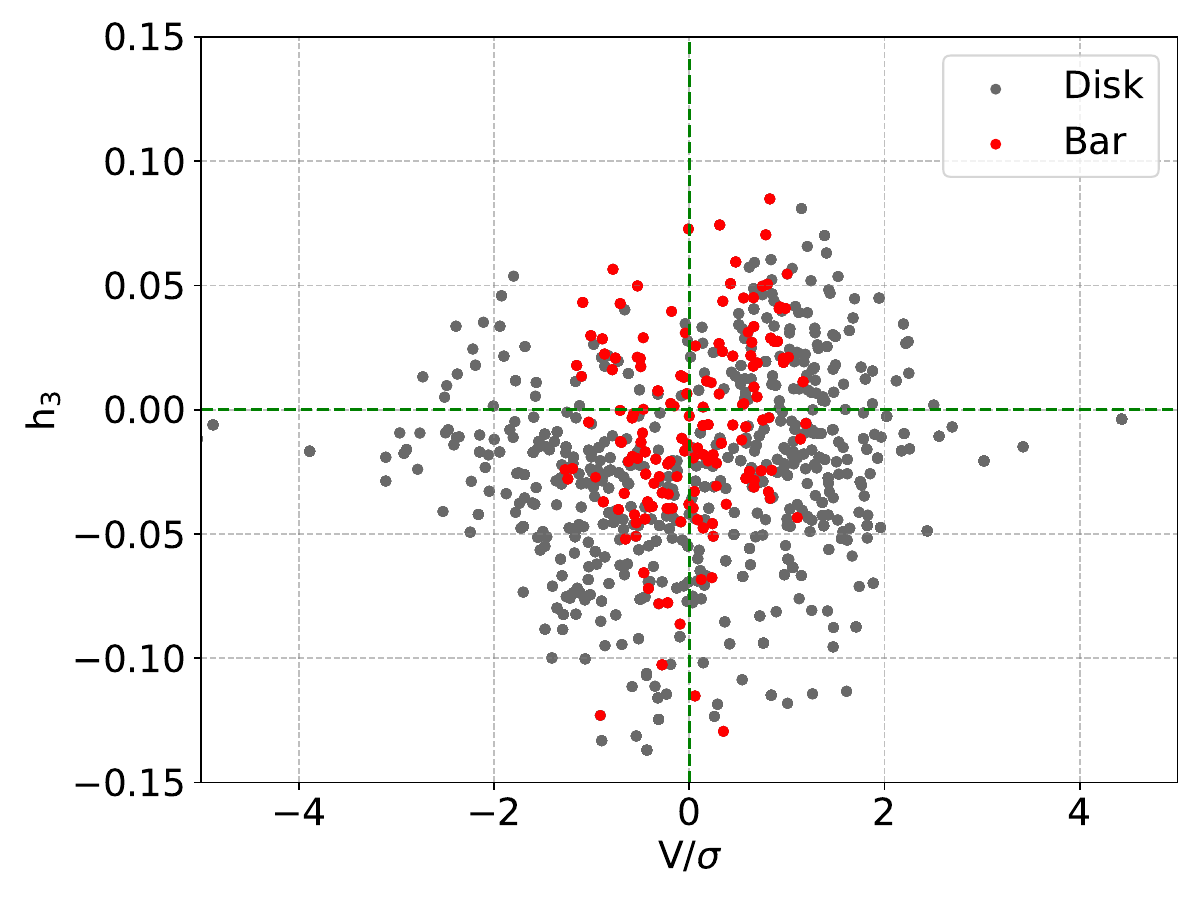}
    \caption{The points show the measured V/$\sigma$ and h$_3$ of each Voronoi bin for the inner region of the Cartwheel. The grey dots denote the bins that are part of the disk, while the bins belonging to the bar (i.e., those within the dotted ellipse shown in Figure \ref{v_by_sigma_bar}) are shown in red.}
    \label{v_by_sigma_bar_ext}
\end{figure}

\begin{figure*}
    \centering
    \includegraphics[width=6.5in]{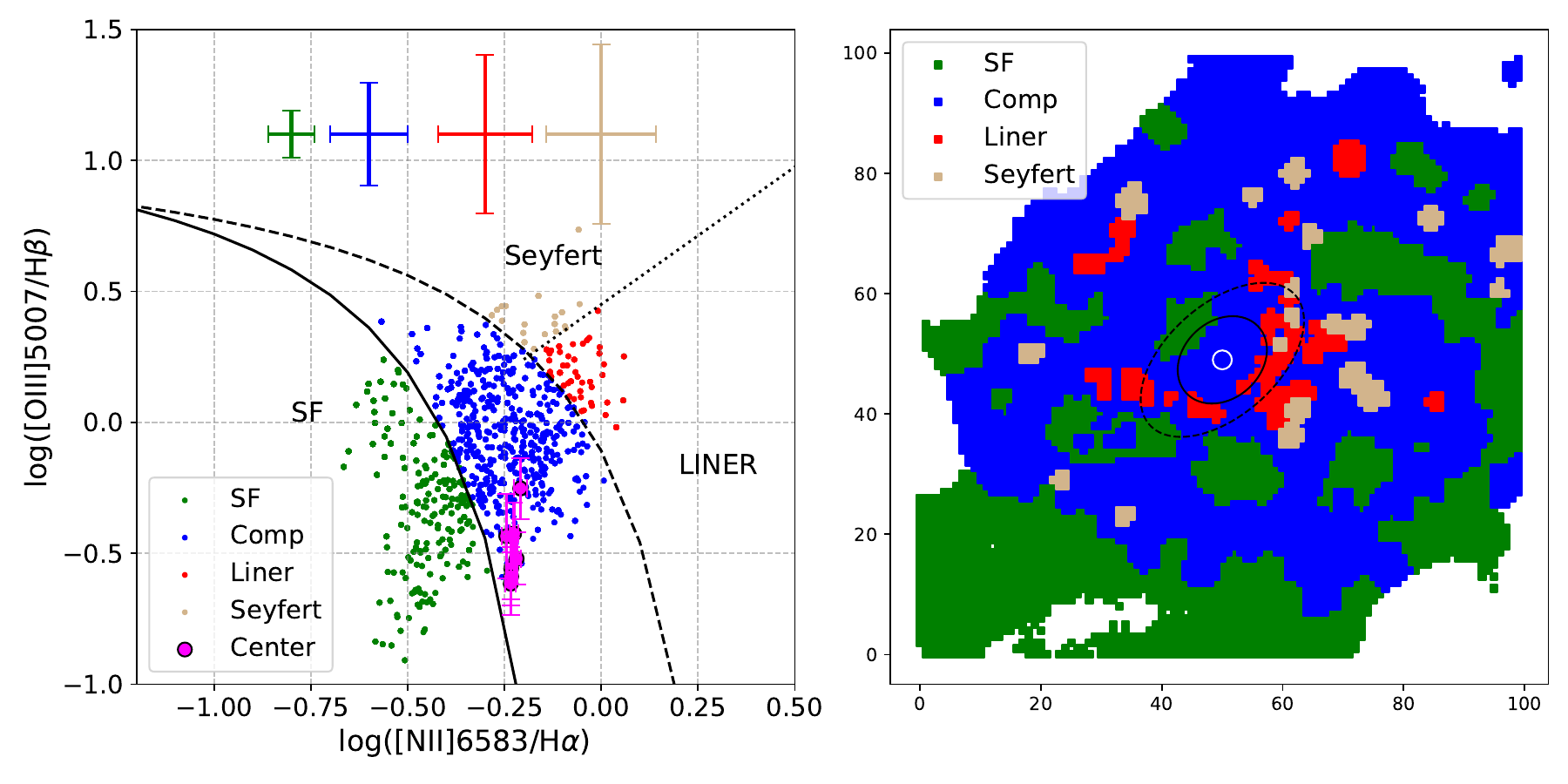}
    \caption{The [NII] BPT diagram of the inner region of the Cartwheel. Each point plotted on the left panel corresponds to the MUSE data cube's single spaxel. Four different colours signify regions of four different ionisation origins of the conventional BPT diagram, i.e., green is for star-forming (SF); blue is for composite; red is for LINER; grey is for Seyfert. The magenta points represent spaxels inside the small white circle shown on the right panel at the Cartwheel center. The error bars shown in different colors (left panel) signify the average uncertainty associated with the measurements of each respective ionisation type. The values and the associated uncertainties for each spaxel belonging to the central white circle (shown on the right panel) are displayed in magenta. The spatial distribution of these points on the Cartwheel is shown in the same corresponding colour on the right panel. The characteristic lines of the BPT diagram shown on the left are taken from \citet{kauffmann2003} (solid), \citet{kewley2001} (dashed), and \citet{schawinski2007} (dotted). We have also shown the extent of the bar and bulge on the right panel, the same as Figure \ref{v_by_sigma_bar}.}
    \label{bpt}
\end{figure*}

\section{Analysis of MUSE Cube}
\label{s_analysis}
We have used the Galaxy IFU Spectroscopy Tool (GIST) pipeline (version - V3.0.2) developed by \citet{bittner2019} for analysing the MUSE data cube. The pipeline is specifically designed to analyse reduced IFS data and has been successfully tested on several IFUs such as MUSE, PPAK (Calar Alto Legacy Integral Field spectroscopy Area - CALIFA), and Spectrograph for INtegral Field Observations in the Near Infrared (SINFONI) \citep{bittner2021,yadav2021,bittner2020,gadotti2020,neumann2020,eisen2003}. We have used the Penalised Pixel-Fitting module (pPXF) \citep{cappellari2004,cappellari2017} of the pipeline to study the stellar kinematics of the galaxy. To apply the pPXF routine, we spatially binned the data using the Voronoi binning algorithm \citep{cappellari2003}. We fixed the targeted S/N at 30 for the binning, allowing us to retain a good spatial resolution of the binned data. The S/N of each spaxel is calculated within the wavelength range 4800 - 5800~\AA~while performing the Voronoi binning. The spaxels with S/N greater than 30 remain unbinned, while those with S/N less than 3 are excluded and not considered in the binning. The integrated spectrum of each binned unit is then fitted with the MILES single stellar population (SSP) model spectra \citep{vazdekis2010} combining with a line-of-sight velocity distribution (LOSVD) of the stellar populations. The LOSVD function is composed of four velocity parameters, i.e., line-of-sight velocity (V), velocity dispersion ($\sigma$), and h$_3$ and h$_4$ Gauss-Hermite moments. The fitting is performed for the continuum of the spectra between the wavelength range 4800 - 5800~\AA. We adopted this wavelength range to avoid contamination from the redder wavelength part as suggested by \citet{bittner2021,bittner2019}. We have masked several prominent spectral lines falling within the considered wavelength range while performing the fit. We also considered an 8th-order multiplicative Legendre polynomial during the fitting to account for small deviations between the model and the observed continuum shape. The best fit is obtained with a least-square minimisation technique. Following the fitting using pPXF, we obtained the values of the four parameters (V, $\sigma$, h$_3$, h$_4$) for each bin. With the best-fit values, we constructed the spatial map of V, $\sigma$, V/$\sigma$, and h$_3$ for the inner part of the galaxy as shown in Figure \ref{v_by_sigma_bar}. As we have studied only the inner part of the galaxy (Figure \ref{cartwheel}) in this work, hence by noting 'galaxy' we mean 'the inner region' in the rest of our text.

In the LOS velocity map (Figure \ref{v_by_sigma_bar}, top-left), we noticed that the north-western region of the galaxy has a blue-shifted velocity, whereas the south-eastern part is red-shifted. The velocity shown in the figure is measured in the rest frame of the galaxy. In the top-right panel of Figure \ref{v_by_sigma_bar}, we have displayed the velocity dispersion map of the galaxy. The dispersion values mostly range between 40 - 80 km/s. In the southeastern part, in between the ring and the central disk, we noticed a region with high dispersion velocity (red bins). The average velocity dispersion in the eastern half is noticed to be relatively higher compared to the western half of the galaxy. Using these V and $\sigma$ maps, we have produced the ratio of LOS velocity and velocity dispersion (i.e., V/$\sigma$) map of the galaxy (bottom-left). The values of V/$\sigma$ mostly lie in the range $\sim$ $-2$ to $+2$, whereas only for the bar region it ranges between $\sim$ $-1$ and $+1$. In the bottom-right panel of Figure \ref{v_by_sigma_bar}, we have shown the h$_3$ velocity moment of the galaxy. We found the values of h$_3$ mostly have a range between $-$0.15 and 0.10 with a slightly lopsided distribution in the central region with respect to the zero velocity line seen in the LOS velocity map. For each kinematic parameter, here we note the typical formal error that arises from the pPXF fitting. The error in the estimated LOS velocity ranges between $\sim$ 5 - 12 km/s, whereas the mean error in the velocity dispersion is around 12 km/s. The error in h$_3$ ranges between $\sim$ 0.05 - 0.18 with a median value $\sim$0.12. For spaxels belonging to the bar, the error in h$_3$ is relatively smaller with a range of $\sim$ 0.05 - 0.11 and median value $\sim$ 0.08. We note here that the formal errors in the derived V, $\sigma$, and $h_3$ values in Cartwheel are slightly higher compared to the nearby systems which have a better sampling on the physical scales with the MUSE.

\section{Results} 
\label{s_result}
\subsection{Kinematic signatures of Bar}
The LOS stellar velocity map of the Cartwheel exhibits a rotating disk in the inner part (Figure \ref{v_by_sigma_bar}, top-left). We noticed a smooth velocity gradient on either side of the rotation axis, which signifies that the central disk and the prominent ring structure are kinematically coupled. The kinematic maps derived from the MUSE data in this study show promising features of an existing bar. Using simulations \citet{bureau2005} discussed multiple kinematic features of a bar in an edge-on disk galaxy. As the disk of the Cartwheel is moderately inclined (\textit{i}$\sim$50 degree; \citet{amram1998}), we examined such features for the Cartwheel in our study. We have used the measured bar semi-major (a), semi-minor (b) axes, and position angle from \citet{barway2020} (as discussed in \S\ref{s_bar_property}) and shown the bar in Figure \ref{v_by_sigma_bar} (top-left) along with the pseudo-bulge. Considering a rectangular pseudo-slit (as shown on the intensity map in Figure \ref{all_profile_bar}), we derived profiles for intensity, LOS velocity, and velocity dispersion along the slit length. The width of the pseudo slit is taken as 2b $\sim$ 18 pixels (i.e., two times the size of the bar's minor axis) and the length is fixed up to the extent of the inner disk. We used a step size of 2 pixels along the slit to define multiple rectangular apertures as shown in the figure. Each rectangular aperture has a size of 2$\times$18 pixels. We estimated the average value of intensity, LOS velocity, and velocity dispersion in all these apertures and plotted along the length of the slit in Figure \ref{all_profile_bar}. The vertical black dashed lines on the right panel mark the extent of the bar on either side, whereas the central line denotes the position of the minor axis. 

We noticed an exponential intensity profile within the bar length with a plateau outside. The LOS velocity profile shows a sharp rise within the bar region where intensity increases rapidly. Both these features are described as the signature of a bar in the simulation by \citet{bureau2005}. The velocity dispersion profile shows a flat profile in the centre with a secondary peak on one side of the bar end. The simulation by \citet{bureau2005} showed a central dip in the velocity dispersion profile if there is no bar in the galaxy. A galaxy with a bar shows a broader peak of different natures depending on the strength and viewing angle. The flatter profile (with a weak peak) noticed in Cartwheel signifies that the galaxy does not host only a disk but also includes an additional dynamical structure that gives rise to the observed velocity dispersion profile. This could also signify the presence of a bar. Therefore, the kinematic features combined with the intensity profile presented in our study support the existence of a bar in this collisional ring galaxy. We note here that a recent work by \citet{iannuzzi2015}, which utilises an upgraded set of dynamical simulations of idealised disk galaxies with different initial gas fractions, attempted to generate 2D kinematic features in LOS velocity, $\sigma$, h$_3$, and h$_4$ of boxy/peanut structures. Their results show similar features in the 2D kinematic maps as discussed through the 1D profiles in \citet{bureau2005}. This further strengthens our interpretation of the presence of a bar in the Cartwheel made based on the 1D simulated profiles reported in \citet{bureau2005}.

To support the bar existence with more kinematic evidence, we used the V/$\sigma$ and $h_3$ maps shown in Figure \ref{v_by_sigma_bar} and plotted their values for all the constructed bins of the galaxy in Figure \ref{v_by_sigma_bar_ext}. Each point in Figure \ref{v_by_sigma_bar_ext} corresponds to the measured values of V/$\sigma$ and h$_3$ for a particular Voronoi bin. Several studies have conveyed that the near-circular stellar orbits in a stable rotating disk show anti-correlation between LOS velocity and h$_3$ moments, whereas a correlation between V/$\sigma$ - h$_3$ points to eccentric stellar orbits, which are unlikely for stable disk \citep{bender1994,bureau2005,gadotti2015,bittner2019,gadotti2020}. A clear correlation between V and $h_3$ is reported by \citet{gadotti2020} for most of the galaxies belonging to the Time Inference with MUSE in Extragalactic Rings (TIMER) project \citep{gadotti2019}, whereas in some cases the correlation is less clear. As each of the TIMER galaxies is known to have a bar, the samples that show less clear correlation might be due to several factors; like the galaxy inclination, dust extinction, or poorer spatial resolution in physical scales \citep{gadotti2020}. In the case of Cartwheel, though the V/$\sigma$ - h$_3$ values do not show a tight correlation or anti-correlation for the entire galaxy, we noticed the bar region alone (red points in Figure \ref{v_by_sigma_bar_ext}) to show a better trend of correlation. We estimated the value of the Spearman correlation coefficient for the bar (red points) and the inner region outside the bar (grey points) and found the values as 0.25 and 0.21 respectively, suggesting weak correlation in both cases. Considering a moderately higher inclination angle and a relatively larger distance of the galaxy, the weak correlation seen in the bar region further strengthens the existence of a stellar bar in Cartwheel. The nature of correlation noticed for the inner part of Cartwheel outside the bar region (in Figure \ref{v_by_sigma_bar_ext}) therefore hints towards a post-collisional disk with less-ordered stellar orbits. The impact of the collision might have produced this anomaly. Besides showing a weak correlation for the bar region in the Cartwheel, the V/$\sigma$ values also match with the typical value of a bar as reported in \citet{gurou2016}. Hence, apart from hosting a disturbed disk, our results present more kinematic evidence for the presence of a bar as well in the Cartwheel.

\begin{figure*}
    \centering
    \includegraphics[width=6.5in]{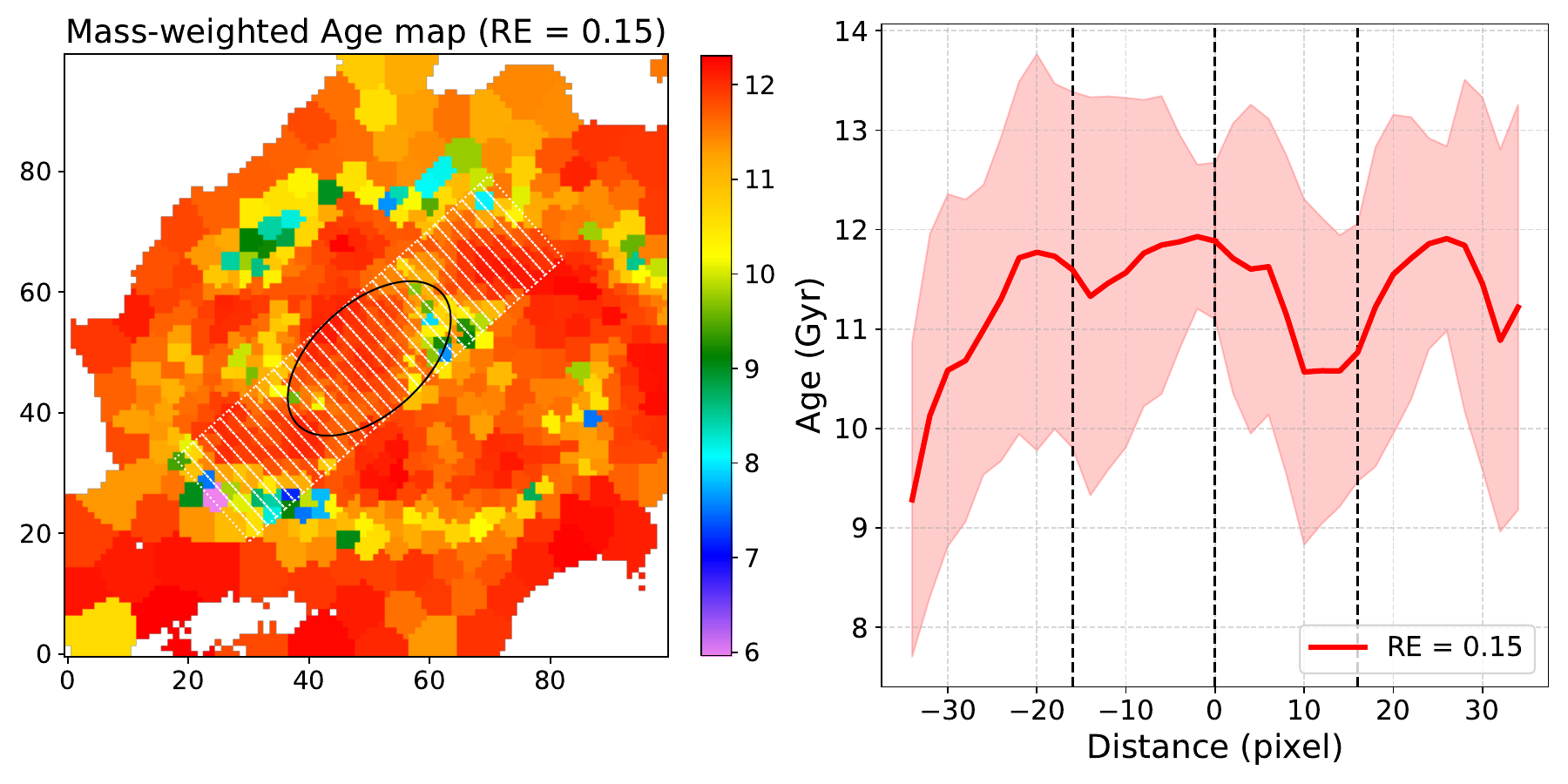}
    \caption{\textbf{Left:} The spatial mass-weighted age map for the inner region of the Cartwheel as derived from binned MUSE cube (for REGUL\_ERR (RE) = 0.15). The extent of the bar is shown with a black ellipse. \textbf{Right:} The age profile is plotted along the rectangular slit (dotted white lines) shown on the left panel. The vertical black dashed line at the centre signifies the position of the minor axis whereas those on either side highlight the bar end. The red shaded region shows the mean difference of the derived age values as estimated from four different regularisation (i.e., RE = 0.15, 0.45, 1.0, and 5.0). The value of the mean age difference for the entire profile is $\sim$ 1.5 Gyr. The age of the stellar populations in the bar is constrained within $\sim$ 9 - 13 Gyr.}

    \label{age_mass}
\end{figure*}

\subsection{Nature of ionised gas}
 \citet{barway2020} reported the detection of an unresolved point source in the NIR $K_s$ band image of the Cartwheel using the GALFIT \citep{peng2002} two-dimensional decomposition code. The presence of an unresolved point source in the NIR $K_s$ band has motivated us to investigate the possible presence of a low luminosity active galactic nuclei (AGN) in the galaxy. The pipeline GIST uses the Gandalf \citep{sarzi2006,falcon2006} procedure to measure the emission-line fluxes. We used the emission line flux map derived from the MUSE data cube to plot the log([OIII]5007/H$\beta$) vs log([NII]6583/H$\alpha$) Baldwin, Phillips \& Terlevich (BPT) diagram as shown in Figure \ref{bpt}. The ratio of these emission lines is measured for each bin and shown on the left panel of the figure. Using the amplitude-to-noise ratio of each emission line from the GIST output, we also estimated average uncertainties associated with each of the four different ionisation types and showed them in the same figure. The characteristic lines that demarcate different natures of ionisation of the BPT plot shown in the Figure \ref{bpt} are taken from the studies of \citet{kewley2001,kauffmann2003,schawinski2007}. On the right panel of the same figure, we have shown the corresponding spatial map of the BPT diagram where each colour signifies the same ionisation nature as mentioned on the left panel. A similar representation of galaxy's BPT map using MUSE data has also been shown earlier by \citet{gadotti2019,lagos2022}. The location of the central spaxels (magenta points on the left panel), those contained within the white circle (right panel) of size $\sim$ MUSE seeing FWHM in the BPT map, signifies that the ionisation in the centre of the galaxy is related to star formation and the presence of AGN in the system is unlikely. However, \citet{struck1996} using the Hubble Space Telescope (HST) WFPC2 imaging in $F450W$ and $F814W$ bands have shown a network of obscuring dust lanes in the central region of the Cartwheel. It is possible that any emission from the accretion disk is not directly detected due to the absorption and/or scattering along the observer's line of sight. The new observations of the Cartwheel from the NIRCAM and NIRSpec instruments onboard the James Webb Space Telescope (JWST) will help to probe the nature of the central point source better to clarify the possibility of an AGN in the Cartwheel. We plan to carry out a follow-up study to understand the NIR characteristics of the central region using the higher resolution JWST observations.

\subsection{Star formation history}
The presence of a stellar bar and a pseudo-bulge that survived a drop-through collision most likely belonged to the pre-collisional progenitor of the Cartwheel. A detailed investigation of the stellar population age can give us more insight into the inner region of Cartwheel. Using the star formation history (SFH) module of GIST, we extracted the mass-weighted age map of the galaxy and showed it in Figure \ref{age_mass}. The SFH module applies regularised pPXF technique \citep{cappellari2003} to fit the emission-line subtracted observed continuum with the model templates. We fixed the input parameter REGUL\_ERR = 0.15 (where Regularisation = 1 / REGUL\_ERR) as suggested in \citet{bittner2020,camila2023} for constraining the strength of regularisation. For the fitting, we limited the wavelength range between 4800 - 6800 ~\AA~. Several emission lines are masked using the default list supplied in GIST while performing the fitting \citep{bittner2019}. We also incorporated an 8th-order multiplicative Legendre polynomial to compensate for the deviation of the continuum between observed and model spectra. The mass-weighted age map shown in Figure \ref{age_mass} highlights that the stellar populations in the central region are relatively older (with age $\sim$ 12 Gyr) than those around it. We placed a pseudo-slit (same as Figure \ref{all_profile_bar}) on the age map and derived the average mass-weighted age profile along the slit length. The older age of the stellar population in the inner region, particularly within the bar, confirms that the structure has a pre-collisional origin and it has survived after the collision. As the stellar population age derived using GIST's SFH module is sensitive to the choice of REGUL\_ERR parameter \citep{bittner2019,camila2023}, we generated the age map for another three different values of regularisation (i.e., REGUL\_ERR = 0.45, 1.0, 5.0). We calculated the mean variation of the age values in the age profile and showed it as a shaded region in Figure \ref{age_mass}. The mean age difference calculated for the entire age profile is found to be $\sim$ 1.5 Gyr. The mass-weighted age of the stellar populations in the bar is constrained within $\sim$ 9 - 13 Gyr, which signifies a pre-collisional origin of the bar in Cartwheel.

We further explored the gas-phase metallicity in the inner region of the Cartwheel. \citet{pettini2004} modelled an empirical relation to estimating oxygen abundance from the measured value of [NII]6583/H$\alpha$ emission line ratio for distant galaxies. To understand the gas phase metallicity, we used the [NII]6583/H$\alpha$ measurements from MUSE data and estimated the 12 + log(O/H) value for each spaxel using equation 1 of \citet{pettini2004}. We found that the gas in the central region of Cartwheel is relatively metal-rich, with 12 + log(O/H) values higher than 9.0\,. \citet{wolter2019} has also reported a super-solar metallicity for the inner region of the Cartwheel. The presence of metal-rich gas in the central region of the galaxy signifies an early enrichment scenario that must have started before the collision.

\section{Summary}
\label{s_summary}
In this work, we have used the MUSE IFU data to study stellar kinematics, SFH, and the nature of ionised gas in the inner region of the Cartwheel, which is an ideal laboratory to test the pre- and post-collisional structure. Using the GIST pipeline, we have produced stellar LOS velocity, velocity dispersion, h$_3$ velocity moment, mass-weighted age, and emission line maps for the inner region of the Cartwheel. The main results of the study are summarised below -

\begin{itemize}
\item The intensity, LOS velocity, and velocity dispersion profiles of the Cartwheel individually show characteristic signatures of a bar as reported in the simulation by \citet{bureau2005}.

\item We found a trend of weak correlation between V/$\sigma$ and h$_3$ for the bar region, which further supports the presence of a pre-discovered bar in the galaxy. The kinematic signature of the bar signifies that the past encounter of the Cartwheel, which resulted in the formation of a ring structure in the system, has not affected the stellar bar to the extent of destroying it.

\item For the inner part of the Cartwheel outside the bar region, we noticed no specific trend of anti-correlation between V/$\sigma$ and h$_3$. It indicates that the collision has affected the stable ordered motion of stars in the inner disk of the galaxy.

\item We produced a BPT map and showed that the ionisation in the galaxy's central region is possibly due to star formation, and the presence of an AGN is unlikely. However, given the dust in the inner region, the higher resolution JWST NIR observations carry a better potential to probe the nature of the central source which we plan to do in the future.

\item The stellar populations within the bar are found to be older (with mass-weighted age constrained between $\sim$ 9 and 13 Gyr). Such older populations in the bar support its pre-collisional existence, which signifies that the bar was already in form when the Cartwheel underwent the collision.

\end{itemize}

In this work, we have addressed whether the bar in Cartwheel was pre-existing or formed after the collision by studying the stellar populations in the bar region, using integral field spectroscopy data from the MUSE. Our finding suggests that the central structures existed before the collision, putting impetus to include inner regions such as bars in future numerical and theoretical studies of collisional ring galaxies. Given the amount of dusty structure, the JWST in the near-IR wavelengths would be ideal for studying the inner region. Thus, collisional ring galaxies provide opportunities to learn a new, less explored aspect of the evolution of the inner region, both observationally and theoretically, which usually survived the collision and remained unaffected by the impact.

\begin{acknowledgements}
This paper has used the observations collected at the European Southern Observatory under ESO program 60.A-9333. This research has made use of the NASA/IPAC Extragalactic Database (NED), which is operated by the Jet Propulsion Laboratory, California Institute of Technology (Caltech), under contract with NASA. We are grateful to Adrian Bittner for his numerous suggestions which helped us to understand the GIST pipeline used in this work. This research made use of Matplotlib \citep{matplotlib2007}, Astropy \citep{astropy2013,astropy2018}, community-developed core Python packages for Astronomy, and SAOImageDS9 \citep{joye2003}. Finally, we thank the referee for valuable suggestions.
\end{acknowledgements}


\begin{thebibliography}{63}
\expandafter\ifx\csname natexlab\endcsname\relax\def\natexlab#1{#1}\fi

\bibitem[{{Aguerri} {et~al.}(2009){Aguerri}, {M{\'e}ndez-Abreu}, \& {Corsini}}]{aguerri2009}
{Aguerri}, J.~A.~L., {M{\'e}ndez-Abreu}, J., \& {Corsini}, E.~M. 2009, \aap, 495, 491

\bibitem[{{Amram} {et~al.}(1998){Amram}, {Mendes de Oliveira}, {Boulesteix}, \& {Balkowski}}]{amram1998}
{Amram}, P., {Mendes de Oliveira}, C., {Boulesteix}, J., \& {Balkowski}, C. 1998, \aap, 330, 881

\bibitem[{{Appleton} \& {Marston}(1997)}]{appleton1997}
{Appleton}, P.~N. \& {Marston}, A.~P. 1997, \aj, 113, 201

\bibitem[{{Astropy Collaboration} {et~al.}(2018){Astropy Collaboration}, {Price-Whelan}, {Sip{\H o}cz}, {G{\"u}nther}, {Lim}, {Crawford}, {Conseil}, {Shupe}, {Craig}, {Dencheva}, {Ginsburg}, {VanderPlas}, {Bradley}, {P{\'e}rez-Su{\'a}rez}, {de Val-Borro}, {Aldcroft}, {Cruz}, {Robitaille}, {Tollerud}, {Ardelean}, {Babej}, {Bach}, {Bachetti}, {Bakanov}, {Bamford}, {Barentsen}, {Barmby}, {Baumbach}, {Berry}, {Biscani}, {Boquien}, {Bostroem}, {Bouma}, {Brammer}, {Bray}, {Breytenbach}, {Buddelmeijer}, {Burke}, {Calderone}, {Cano Rodr{\'{\i}}guez}, {Cara}, {Cardoso}, {Cheedella}, {Copin}, {Corrales}, {Crichton}, {D'Avella}, {Deil}, {Depagne}, {Dietrich}, {Donath}, {Droettboom}, {Earl}, {Erben}, {Fabbro}, {Ferreira}, {Finethy}, {Fox}, {Garrison}, {Gibbons}, {Goldstein}, {Gommers}, {Greco}, {Greenfield}, {Groener}, {Grollier}, {Hagen}, {Hirst}, {Homeier}, {Horton}, {Hosseinzadeh}, {Hu}, {Hunkeler}, {Ivezi{\'c}}, {Jain}, {Jenness}, {Kanarek}, {Kendrew}, {Kern}, {Kerzendorf}, {Khvalko}, {King}, {Kirkby}, {Kulkarni},
  {Kumar}, {Lee}, {Lenz}, {Littlefair}, {Ma}, {Macleod}, {Mastropietro}, {McCully}, {Montagnac}, {Morris}, {Mueller}, {Mumford}, {Muna}, {Murphy}, {Nelson}, {Nguyen}, {Ninan}, {N{\"o}the}, {Ogaz}, {Oh}, {Parejko}, {Parley}, {Pascual}, {Patil}, {Patil}, {Plunkett}, {Prochaska}, {Rastogi}, {Reddy Janga}, {Sabater}, {Sakurikar}, {Seifert}, {Sherbert}, {Sherwood-Taylor}, {Shih}, {Sick}, {Silbiger}, {Singanamalla}, {Singer}, {Sladen}, {Sooley}, {Sornarajah}, {Streicher}, {Teuben}, {Thomas}, {Tremblay}, {Turner}, {Terr{\'o}n}, {van Kerkwijk}, {de la Vega}, {Watkins}, {Weaver}, {Whitmore}, {Woillez}, {Zabalza}, \& {Astropy Contributors}}]{astropy2018}
{Astropy Collaboration}, {Price-Whelan}, A.~M., {Sip{\H o}cz}, B.~M., {et~al.} 2018, \aj, 156, 123

\bibitem[{{Astropy Collaboration} {et~al.}(2013){Astropy Collaboration}, {Robitaille}, {Tollerud}, {Greenfield}, {Droettboom}, {Bray}, {Aldcroft}, {Davis}, {Ginsburg}, {Price-Whelan}, {Kerzendorf}, {Conley}, {Crighton}, {Barbary}, {Muna}, {Ferguson}, {Grollier}, {Parikh}, {Nair}, {Unther}, {Deil}, {Woillez}, {Conseil}, {Kramer}, {Turner}, {Singer}, {Fox}, {Weaver}, {Zabalza}, {Edwards}, {Azalee Bostroem}, {Burke}, {Casey}, {Crawford}, {Dencheva}, {Ely}, {Jenness}, {Labrie}, {Lim}, {Pierfederici}, {Pontzen}, {Ptak}, {Refsdal}, {Servillat}, \& {Streicher}}]{astropy2013}
{Astropy Collaboration}, {Robitaille}, T.~P., {Tollerud}, E.~J., {et~al.} 2013, \aap, 558, A33

\bibitem[{{Athanassoula} {et~al.}(2005){Athanassoula}, {Lambert}, \& {Dehnen}}]{athan2005}
{Athanassoula}, E., {Lambert}, J.~C., \& {Dehnen}, W. 2005, \mnras, 363, 496

\bibitem[{{Athanassoula} {et~al.}(1997){Athanassoula}, {Puerari}, \& {Bosma}}]{athan1997}
{Athanassoula}, E., {Puerari}, I., \& {Bosma}, A. 1997, \mnras, 286, 284

\bibitem[{{Bacon} {et~al.}(2010){Bacon}, {Accardo}, {Adjali}, {Anwand}, {Bauer}, {Biswas}, {Blaizot}, {Boudon}, {Brau-Nogue}, {Brinchmann}, {Caillier}, {Capoani}, {Carollo}, {Contini}, {Couderc}, {Daguis{\'e}}, {Deiries}, {Delabre}, {Dreizler}, {Dubois}, {Dupieux}, {Dupuy}, {Emsellem}, {Fechner}, {Fleischmann}, {Fran{\c{c}}ois}, {Gallou}, {Gharsa}, {Glindemann}, {Gojak}, {Guiderdoni}, {Hansali}, {Hahn}, {Jarno}, {Kelz}, {Koehler}, {Kosmalski}, {Laurent}, {Le Floch}, {Lilly}, {Lizon}, {Loupias}, {Manescau}, {Monstein}, {Nicklas}, {Olaya}, {Pares}, {Pasquini}, {P{\'e}contal-Rousset}, {Pell{\'o}}, {Petit}, {Popow}, {Reiss}, {Remillieux}, {Renault}, {Roth}, {Rupprecht}, {Serre}, {Schaye}, {Soucail}, {Steinmetz}, {Streicher}, {Stuik}, {Valentin}, {Vernet}, {Weilbacher}, {Wisotzki}, \& {Yerle}}]{bacon2010}
{Bacon}, R., {Accardo}, M., {Adjali}, L., {et~al.} 2010, in Society of Photo-Optical Instrumentation Engineers (SPIE) Conference Series, Vol. 7735, Ground-based and Airborne Instrumentation for Astronomy III, ed. I.~S. {McLean}, S.~K. {Ramsay}, \& H.~{Takami}, 773508

\bibitem[{{Barway} {et~al.}(2020){Barway}, {Mayya}, \& {Robleto-Or{\'u}s}}]{barway2020}
{Barway}, S., {Mayya}, Y.~D., \& {Robleto-Or{\'u}s}, A. 2020, \mnras, 497, 44

\bibitem[{{Barway} \& {Saha}(2020)}]{barway2020a}
{Barway}, S. \& {Saha}, K. 2020, \mnras, 495, 4548

\bibitem[{{Barway} {et~al.}(2011){Barway}, {Wadadekar}, \& {Kembhavi}}]{barway2011}
{Barway}, S., {Wadadekar}, Y., \& {Kembhavi}, A.~K. 2011, \mnras, 410, L18

\bibitem[{{Bender} {et~al.}(1994){Bender}, {Saglia}, \& {Gerhard}}]{bender1994}
{Bender}, R., {Saglia}, R.~P., \& {Gerhard}, O.~E. 1994, \mnras, 269, 785

\bibitem[{{Bittner} {et~al.}(2021){Bittner}, {de Lorenzo-C{\'a}ceres}, {Gadotti}, {S{\'a}nchez-Bl{\'a}zquez}, {Neumann}, {Coelho}, {Falc{\'o}n-Barroso}, {Fragkoudi}, {Kim}, {Mart{\'\i}n-Navarro}, {M{\'e}ndez-Abreu}, {P{\'e}rez}, {Querejeta}, \& {van de Ven}}]{bittner2021}
{Bittner}, A., {de Lorenzo-C{\'a}ceres}, A., {Gadotti}, D.~A., {et~al.} 2021, \aap, 646, A42

\bibitem[{{Bittner} {et~al.}(2019){Bittner}, {Falc{\'o}n-Barroso}, {Nedelchev}, {Dorta}, {Gadotti}, {Sarzi}, {Molaeinezhad}, {Iodice}, {Rosado-Belza}, {de Lorenzo-C{\'a}ceres}, {Fragkoudi}, {Gal{\'a}n-de Anta}, {Husemann}, {M{\'e}ndez-Abreu}, {Neumann}, {Pinna}, {Querejeta}, {S{\'a}nchez-Bl{\'a}zquez}, \& {Seidel}}]{bittner2019}
{Bittner}, A., {Falc{\'o}n-Barroso}, J., {Nedelchev}, B., {et~al.} 2019, \aap, 628, A117

\bibitem[{{Bittner} {et~al.}(2020){Bittner}, {S{\'a}nchez-Bl{\'a}zquez}, {Gadotti}, {Neumann}, {Fragkoudi}, {Coelho}, {de Lorenzo-C{\'a}ceres}, {Falc{\'o}n-Barroso}, {Kim}, {Leaman}, {Mart{\'\i}n-Navarro}, {M{\'e}ndez-Abreu}, {P{\'e}rez}, {Querejeta}, {Seidel}, \& {van de Ven}}]{bittner2020}
{Bittner}, A., {S{\'a}nchez-Bl{\'a}zquez}, P., {Gadotti}, D.~A., {et~al.} 2020, \aap, 643, A65

\bibitem[{{Bournaud} \& {Combes}(2002)}]{bournaud2002}
{Bournaud}, F. \& {Combes}, F. 2002, \aap, 392, 83

\bibitem[{{Bureau} \& {Athanassoula}(2005)}]{bureau2005}
{Bureau}, M. \& {Athanassoula}, E. 2005, \apj, 626, 159

\bibitem[{{Cappellari}(2017)}]{cappellari2017}
{Cappellari}, M. 2017, \mnras, 466, 798

\bibitem[{{Cappellari} \& {Copin}(2003)}]{cappellari2003}
{Cappellari}, M. \& {Copin}, Y. 2003, \mnras, 342, 345

\bibitem[{{Cappellari} \& {Emsellem}(2004)}]{cappellari2004}
{Cappellari}, M. \& {Emsellem}, E. 2004, \pasp, 116, 138

\bibitem[{{Cardelli} {et~al.}(1989){Cardelli}, {Clayton}, \& {Mathis}}]{cardelli1989}
{Cardelli}, J.~A., {Clayton}, G.~C., \& {Mathis}, J.~S. 1989, \apj, 345, 245

\bibitem[{{Charmandaris} \& {Mirabel}(1997)}]{charmandaris1997}
{Charmandaris}, V. \& {Mirabel}, F. 1997, in Joint European and National Astronomical Meeting, ed. J.~D. {Hadjidemetrioy} \& J.~H. {Seiradakis}, 185

\bibitem[{{de S{\'a}-Freitas} {et~al.}(2023){de S{\'a}-Freitas}, {Fragkoudi}, {Gadotti}, {Falc{\'o}n-Barroso}, {Bittner}, {S{\'a}nchez-Bl{\'a}zquez}, {van de Ven}, {Bieri}, {Coccato}, {Coelho}, {Fahrion}, {Gon{\c{c}}alves}, {Kim}, {de Lorenzo-C{\'a}ceres}, {Martig}, {Mart{\'\i}n-Navarro}, {Mendez-Abreu}, {Neumann}, \& {Querejeta}}]{camila2023}
{de S{\'a}-Freitas}, C., {Fragkoudi}, F., {Gadotti}, D.~A., {et~al.} 2023, \aap, 671, A8

\bibitem[{{Donohoe-Keyes} {et~al.}(2019){Donohoe-Keyes}, {Martig}, {James}, \& {Kraljic}}]{dono2019}
{Donohoe-Keyes}, C.~E., {Martig}, M., {James}, P.~A., \& {Kraljic}, K. 2019, \mnras, 489, 4992

\bibitem[{{Eisenhauer} {et~al.}(2003){Eisenhauer}, {Abuter}, {Bickert}, {Biancat-Marchet}, {Bonnet}, {Brynnel}, {Conzelmann}, {Delabre}, {Donaldson}, {Farinato}, {Fedrigo}, {Genzel}, {Hubin}, {Iserlohe}, {Kasper}, {Kissler-Patig}, {Monnet}, {Roehrle}, {Schreiber}, {Stroebele}, {Tecza}, {Thatte}, \& {Weisz}}]{eisen2003}
{Eisenhauer}, F., {Abuter}, R., {Bickert}, K., {et~al.} 2003, in Society of Photo-Optical Instrumentation Engineers (SPIE) Conference Series, Vol. 4841, Instrument Design and Performance for Optical/Infrared Ground-based Telescopes, ed. M.~{Iye} \& A.~F.~M. {Moorwood}, 1548--1561

\bibitem[{{Ellison} {et~al.}(2011){Ellison}, {Nair}, {Patton}, {Scudder}, {Mendel}, \& {Simard}}]{ellison2011}
{Ellison}, S.~L., {Nair}, P., {Patton}, D.~R., {et~al.} 2011, \mnras, 416, 2182

\bibitem[{{Eskridge} {et~al.}(2000){Eskridge}, {Frogel}, {Pogge}, {Quillen}, {Davies}, {DePoy}, {Houdashelt}, {Kuchinski}, {Ram{\'\i}rez}, {Sellgren}, {Terndrup}, \& {Tiede}}]{eskridge2000}
{Eskridge}, P.~B., {Frogel}, J.~A., {Pogge}, R.~W., {et~al.} 2000, \aj, 119, 536

\bibitem[{{Falc{\'o}n-Barroso} {et~al.}(2006){Falc{\'o}n-Barroso}, {Bacon}, {Bureau}, {Cappellari}, {Davies}, {de Zeeuw}, {Emsellem}, {Fathi}, {Krajnovi{\'c}}, {Kuntschner}, {McDermid}, {Peletier}, \& {Sarzi}}]{falcon2006}
{Falc{\'o}n-Barroso}, J., {Bacon}, R., {Bureau}, M., {et~al.} 2006, \mnras, 369, 529

\bibitem[{{Gadotti} {et~al.}(2020){Gadotti}, {Bittner}, {Falc{\'o}n-Barroso}, {M{\'e}ndez-Abreu}, {Kim}, {Fragkoudi}, {de Lorenzo-C{\'a}ceres}, {Leaman}, {Neumann}, {Querejeta}, {S{\'a}nchez-Bl{\'a}zquez}, {Martig}, {Mart{\'\i}n-Navarro}, {P{\'e}rez}, {Seidel}, \& {van de Ven}}]{gadotti2020}
{Gadotti}, D.~A., {Bittner}, A., {Falc{\'o}n-Barroso}, J., {et~al.} 2020, \aap, 643, A14

\bibitem[{{Gadotti} {et~al.}(2019){Gadotti}, {S{\'a}nchez-Bl{\'a}zquez}, {Falc{\'o}n-Barroso}, {Husemann}, {Seidel}, {P{\'e}rez}, {de Lorenzo-C{\'a}ceres}, {Martinez-Valpuesta}, {Fragkoudi}, {Leung}, {van de Ven}, {Leaman}, {Coelho}, {Martig}, {Kim}, {Neumann}, \& {Querejeta}}]{gadotti2019}
{Gadotti}, D.~A., {S{\'a}nchez-Bl{\'a}zquez}, P., {Falc{\'o}n-Barroso}, J., {et~al.} 2019, \mnras, 482, 506

\bibitem[{{Gadotti} {et~al.}(2015){Gadotti}, {Seidel}, {S{\'a}nchez-Bl{\'a}zquez}, {Falc{\'o}n-Barroso}, {Husemann}, {Coelho}, \& {P{\'e}rez}}]{gadotti2015}
{Gadotti}, D.~A., {Seidel}, M.~K., {S{\'a}nchez-Bl{\'a}zquez}, P., {et~al.} 2015, \aap, 584, A90

\bibitem[{{Gu{\'e}rou} {et~al.}(2016){Gu{\'e}rou}, {Emsellem}, {Krajnovi{\'c}}, {McDermid}, {Contini}, \& {Weilbacher}}]{gurou2016}
{Gu{\'e}rou}, A., {Emsellem}, E., {Krajnovi{\'c}}, D., {et~al.} 2016, \aap, 591, A143

\bibitem[{{Higdon}(1996)}]{higdon1996}
{Higdon}, J.~L. 1996, \apj, 467, 241

\bibitem[{{Higdon} {et~al.}(2015){Higdon}, {Higdon}, {Mart{\'\i}n Ruiz}, \& {Rand}}]{higdon2015}
{Higdon}, J.~L., {Higdon}, S. J.~U., {Mart{\'\i}n Ruiz}, S., \& {Rand}, R.~J. 2015, \apjl, 814, L1

\bibitem[{Hunter(2007)}]{matplotlib2007}
Hunter, J.~D. 2007, Computing In Science \& Engineering, 9, 90

\bibitem[{{Iannuzzi} \& {Athanassoula}(2015)}]{iannuzzi2015}
{Iannuzzi}, F. \& {Athanassoula}, E. 2015, \mnras, 450, 2514

\bibitem[{{Joye} \& {Mandel}(2003)}]{joye2003}
{Joye}, W.~A. \& {Mandel}, E. 2003, in Astronomical Society of the Pacific Conference Series, Vol. 295, Astronomical Data Analysis Software and Systems XII, ed. H.~E. {Payne}, R.~I. {Jedrzejewski}, \& R.~N. {Hook}, 489

\bibitem[{{Kauffmann} {et~al.}(2003){Kauffmann}, {Heckman}, {Tremonti}, {Brinchmann}, {Charlot}, {White}, {Ridgway}, {Brinkmann}, {Fukugita}, {Hall}, {Ivezi{\'c}}, {Richards}, \& {Schneider}}]{kauffmann2003}
{Kauffmann}, G., {Heckman}, T.~M., {Tremonti}, C., {et~al.} 2003, \mnras, 346, 1055

\bibitem[{{Kewley} {et~al.}(2001){Kewley}, {Dopita}, {Sutherland}, {Heisler}, \& {Trevena}}]{kewley2001}
{Kewley}, L.~J., {Dopita}, M.~A., {Sutherland}, R.~S., {Heisler}, C.~A., \& {Trevena}, J. 2001, \apj, 556, 121

\bibitem[{{Kim} {et~al.}(2016){Kim}, {Gadotti}, {Athanassoula}, {Bosma}, {Sheth}, \& {Lee}}]{kim2016}
{Kim}, T., {Gadotti}, D.~A., {Athanassoula}, E., {et~al.} 2016, \mnras, 462, 3430

\bibitem[{{Kormendy} \& {Kennicutt}(2004)}]{kormendy2004}
{Kormendy}, J. \& {Kennicutt}, Robert~C., J. 2004, \araa, 42, 603

\bibitem[{{Kraljic} {et~al.}(2012){Kraljic}, {Bournaud}, \& {Martig}}]{kraljic2012}
{Kraljic}, K., {Bournaud}, F., \& {Martig}, M. 2012, \apj, 757, 60

\bibitem[{{Kruk} {et~al.}(2018){Kruk}, {Lintott}, {Bamford}, {Masters}, {Simmons}, {H{\"a}u{\ss}ler}, {Cardamone}, {Hart}, {Kelvin}, {Schawinski}, {Smethurst}, \& {Vika}}]{kruk2018}
{Kruk}, S.~J., {Lintott}, C.~J., {Bamford}, S.~P., {et~al.} 2018, \mnras, 473, 4731

\bibitem[{{Lagos} {et~al.}(2022){Lagos}, {Loubser}, {Scott}, {O'Sullivan}, {Kolokythas}, {Babul}, {Nigoche-Netro}, {Olivares}, \& {Sengupta}}]{lagos2022}
{Lagos}, P., {Loubser}, S.~I., {Scott}, T.~C., {et~al.} 2022, \mnras, 516, 5487

\bibitem[{{Lynds} \& {Toomre}(1976)}]{lynds1976}
{Lynds}, R. \& {Toomre}, A. 1976, \apj, 209, 382

\bibitem[{{Marcum} {et~al.}(1992){Marcum}, {Appleton}, \& {Higdon}}]{marcum1992}
{Marcum}, P.~M., {Appleton}, P.~N., \& {Higdon}, J.~L. 1992, \apj, 399, 57

\bibitem[{{Marinova} \& {Jogee}(2007)}]{marinova2007}
{Marinova}, I. \& {Jogee}, S. 2007, \apj, 659, 1176

\bibitem[{{Marston} \& {Appleton}(1995)}]{marston1995}
{Marston}, A.~P. \& {Appleton}, P.~N. 1995, \aj, 109, 1002

\bibitem[{{Mayya} {et~al.}(2023){Mayya}, {Plat}, {G{\'o}mez-Gonz{\'a}lez}, {Zaragoza-Cardiel}, {Charlot}, \& {Bruzual}}]{mayya2023}
{Mayya}, Y.~D., {Plat}, A., {G{\'o}mez-Gonz{\'a}lez}, V.~M.~A., {et~al.} 2023, \mnras, 519, 5492

\bibitem[{{Men{\'e}ndez-Delmestre} {et~al.}(2007){Men{\'e}ndez-Delmestre}, {Sheth}, {Schinnerer}, {Jarrett}, \& {Scoville}}]{menendez2007}
{Men{\'e}ndez-Delmestre}, K., {Sheth}, K., {Schinnerer}, E., {Jarrett}, T.~H., \& {Scoville}, N.~Z. 2007, \apj, 657, 790

\bibitem[{{Neumann} {et~al.}(2020){Neumann}, {Fragkoudi}, {P{\'e}rez}, {Gadotti}, {Falc{\'o}n-Barroso}, {S{\'a}nchez-Bl{\'a}zquez}, {Bittner}, {Husemann}, {G{\'o}mez}, {Grand}, {Donohoe-Keyes}, {Kim}, {de Lorenzo-C{\'a}ceres}, {Martig}, {M{\'e}ndez-Abreu}, {Pakmor}, {Seidel}, \& {van de Ven}}]{neumann2020}
{Neumann}, J., {Fragkoudi}, F., {P{\'e}rez}, I., {et~al.} 2020, \aap, 637, A56

\bibitem[{{Peng} {et~al.}(2002){Peng}, {Ho}, {Impey}, \& {Rix}}]{peng2002}
{Peng}, C.~Y., {Ho}, L.~C., {Impey}, C.~D., \& {Rix}, H.-W. 2002, \aj, 124, 266

\bibitem[{{Pettini} \& {Pagel}(2004)}]{pettini2004}
{Pettini}, M. \& {Pagel}, B. E.~J. 2004, \mnras, 348, L59

\bibitem[{{Renaud} {et~al.}(2018){Renaud}, {Athanassoula}, {Amram}, {Bosma}, {Bournaud}, {Duc}, {Epinat}, {Fensch}, {Kraljic}, {Perret}, \& {Struck}}]{renaud2018}
{Renaud}, F., {Athanassoula}, E., {Amram}, P., {et~al.} 2018, \mnras, 473, 585

\bibitem[{{Riad} {et~al.}(2010){Riad}, {Kraan-Korteweg}, \& {Woudt}}]{riad2010}
{Riad}, I.~F., {Kraan-Korteweg}, R.~C., \& {Woudt}, P.~A. 2010, \mnras, 401, 924

\bibitem[{{Sakamoto} {et~al.}(1999){Sakamoto}, {Okumura}, {Ishizuki}, \& {Scoville}}]{sakamoto1999}
{Sakamoto}, K., {Okumura}, S.~K., {Ishizuki}, S., \& {Scoville}, N.~Z. 1999, \apj, 525, 691

\bibitem[{{Sarzi} {et~al.}(2006){Sarzi}, {Falc{\'o}n-Barroso}, {Davies}, {Bacon}, {Bureau}, {Cappellari}, {de Zeeuw}, {Emsellem}, {Fathi}, {Krajnovi{\'c}}, {Kuntschner}, {McDermid}, \& {Peletier}}]{sarzi2006}
{Sarzi}, M., {Falc{\'o}n-Barroso}, J., {Davies}, R.~L., {et~al.} 2006, \mnras, 366, 1151

\bibitem[{{Schawinski} {et~al.}(2007){Schawinski}, {Thomas}, {Sarzi}, {Maraston}, {Kaviraj}, {Joo}, {Yi}, \& {Silk}}]{schawinski2007}
{Schawinski}, K., {Thomas}, D., {Sarzi}, M., {et~al.} 2007, \mnras, 382, 1415

\bibitem[{{Struck} {et~al.}(1996){Struck}, {Appleton}, {Borne}, \& {Lucas}}]{struck1996}
{Struck}, C., {Appleton}, P.~N., {Borne}, K.~D., \& {Lucas}, R.~A. 1996, \aj, 112, 1868

\bibitem[{{Vazdekis} {et~al.}(2010){Vazdekis}, {S{\'a}nchez-Bl{\'a}zquez}, {Falc{\'o}n-Barroso}, {Cenarro}, {Beasley}, {Cardiel}, {Gorgas}, \& {Peletier}}]{vazdekis2010}
{Vazdekis}, A., {S{\'a}nchez-Bl{\'a}zquez}, P., {Falc{\'o}n-Barroso}, J., {et~al.} 2010, \mnras, 404, 1639

\bibitem[{{Wolter} {et~al.}(2019){Wolter}, {Consolandi}, {Longhetti}, {Landoni}, \& {Bianco}}]{wolter2019}
{Wolter}, A., {Consolandi}, G., {Longhetti}, M., {Landoni}, M., \& {Bianco}, A. 2019, IAU Symposium, 346, 297

\bibitem[{{Yadav} {et~al.}(2021){Yadav}, {Das}, {Barway}, \& {Combes}}]{yadav2021}
{Yadav}, J., {Das}, M., {Barway}, S., \& {Combes}, F. 2021, \aap, 651, L9

\bibitem[{{Zaragoza-Cardiel} {et~al.}(2022){Zaragoza-Cardiel}, {G{\'o}mez-Gonz{\'a}lez}, {Mayya}, \& {Ramos-Larios}}]{zaragoza2022}
{Zaragoza-Cardiel}, J., {G{\'o}mez-Gonz{\'a}lez}, V. M.~A., {Mayya}, D., \& {Ramos-Larios}, G. 2022, \mnras, 514, 1689

\end{thebibliography}

\end{document}